# FERMILAB MAIN INJECTOR AND RECYCLER OPERATIONS IN THE MEGAWATT ERA

A.P. Schreckenberger\*, Fermi National Accelerator Laboratory, Batavia, IL, U.S.A.

*Abstract*

Significant upgrades to Fermilab's accelerator complex have accompanied the development of LBNF (the Long Baseline Neutrino Facility) and DUNE (Deep Underground Neutrino Experiment). These improvements will facilitate 1-MW operation of the NuMI (Neutrinos at the Main Injector) beam for the first time this year through changes to the Recycler slip-stacking procedure and shortening of the Main Injector ramp time. The modifications to the Recycler slip-stacking and efforts to reduce the duration of the Main Injector ramp will be discussed. Additionally, details regarding further shortening of the ramp time and the subsequent impacts on future accelerator operations are presented.

## FERMILAB MOTIVATIONS

The Fermilab accelerator complex, illustrated in Fig. 1, delivers protons and consequent collision decay products to various physics experiments that include studies of accelerator stability, beam dynamics, muons, low- and high-energy neutrino interactions, neutrino oscillations, and the fixed-target program.

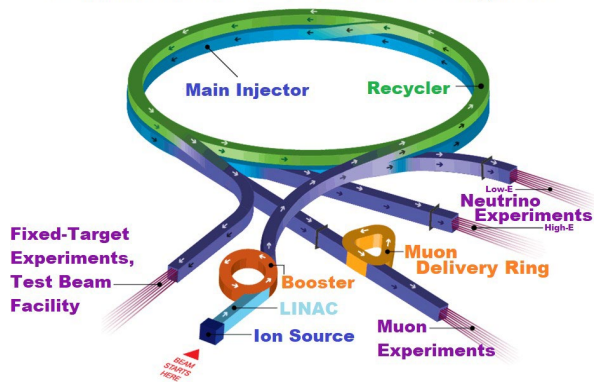

Figure 1: Illustration of the current Fermilab accelerator complex.

Fermilab's highest priority in the coming decade will be the development and construction of LBNF and DUNE [1]. DUNE's large liquid-argon-based detectors will provide unprecedented measurement precision of the neutrino oscillation parameters and sensitivity to CP-violation in the neutrino sector [2], paving a critical path to probing one of the Universe's fundamental questions.

Maximizing the neutrino flux produced by LBNF will improve the physics reach of DUNE. This goal is directly tied to the beam power delivered from the Main Injector (MI) and the efficiency of the accelerator complex. Many components in the proton accelerator chain have been in operation for decades and have exceeded their intended lifetimes.

The Fermilab linear accelerator (LINAC) was commissioned in 1970 and reached an upgraded design energy of 400 MeV in 1993. Construction has already begun on a new superconducting RF LINAC as part of the Proton Improvement Plan II (PIP-II) that will replace the original machine [3]. Likewise, the Booster synchrotron, responsible for accelerating protons to 8 GeV, reached its designed running in 1971. The existing proton source and downstream components operate at 15 Hz, but the PIP-II accelerator will operate at 20 Hz, further pushing the capabilities of the 52-year-old Booster.

The Recycler is an 8-GeV permanent magnet storage ring designed to accumulate antiprotons during Tevatron operations. Presently, the accelerator delivers beam to the Muon Campus and facilitates slip-stacking, which doubles the bunch intensity prior to delivery to the Main Injector. Occupying the same tunnel enclosure as the Recycler, the Main Injector accelerates the input 8-GeV beam to 120 GeV and provides protons to high-energy neutrino and test-beam experiments.

Beam power can be represented as:

$$P = |e| E \frac{N}{T}, \qquad (1)$$

where e is the elementary charge, E is the particle energy, N is the number of particles per pulse, and T is the cycle duration. The Main Injector will continue to deliver protons at 120 GeV, so increases in beam power must be driven by increasing the beam intensity or decreasing the cycle duration. The slip-stacking process preloads an intensity increase into the Main Injector. However, additional increases do not come without incurring risks. Slip-stacking itself; the rise of fast-transverse, convective, and transverse mode coupling instabilities (TMCI); electron cloud accumulation, and space charge tune shifts are a few mechanisms that can yield additional beam losses as beam intensity grows. Recent slip-stacking optimization was an essential step for megawatt and LBNF/DUNE operations.

Decreasing the cycle duration is achievable by reducing the MI ramp time. During nominal NuMI running, this duration is 1.2 s, and megawatt operations are reached via a reduction to 1.067 s. As part of the Accelerator Complex Evolution (ACE) initiative, we aim to reduce the ramp time to 0.65 s to produce a multi-megawatt neutrino beam for LBNF. The advantages and requirements of this approach are discussed in this manuscript.

\* wingmc@fnal.gov

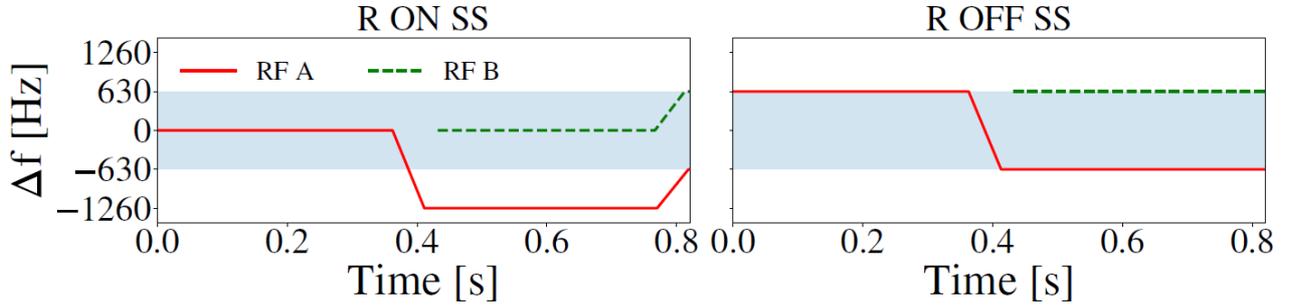

Figure 2: Frequencies of the RF systems used in slip-stacking to compare the two implemented methods in the Fermilab Recycler. *Left:* R ON SS designates the original slip-stacking scheme. *Right:* R OFF SS depicts the Radial Off-Center approach. The shaded regions facilitate quick comparisons of the apertures imposed by the two cases. From Ref. [4].

## SLIP-STACKING IMPROVEMENTS

The Recycler accepts six batches of 84 bunches from the Booster, which encroaches intensity limitations. Slip-stacking in the Recycler was introduced to bypass this Booster limit, facilitating the injection of 12 batches that are subsequently combined into six batches with double the initial intensity.

In the original implementation, slip-stacking worked by injecting six batches at the radial center of the Recycler aperture. The batches were then decelerated by 1260 Hz, clearing the way for six additional batches to be injected at the designed machine frequency/radial center. Due to the momentum difference between the two sets, the decelerated group slips with respect to the on-frequency batches until the groups overlap before extraction to the Main Injector. An additional 630-Hz acceleration is applied just prior to extraction to impose radial symmetry.

With the plan to increase the Booster cycle rate to 20 Hz in the PIP-II era, the separation frequency (given by Eq. 2) increases to 1680 Hz. The larger momentum aperture could produce additional beam losses, and as such, an alternative slip-stacking process was investigated.

$$\Delta f = h_b f_b, \qquad (2)$$

where $h_b$ is the Booster harmonic number (84) and $f_b$ is the Booster cycle rate.

We implemented Radial Off-Center Slip-Stacking to address the momentum aperture issue. In this procedure, the first six batches are injected 630 Hz above the design frequency, placing the protons off center in the machine. The deceleration is once again applied to clear the way for the second set, again injected above the design frequency.

Figure 2 depicts the RF frequencies used by the slip-stacking systems during both the standard, Radial On-Center Slip-Stacking (R ON SS) and new, Radial Off-Center Slip-Stacking (R OFF SS) methods. While running at the 15-Hz cycle rate, it is apparent that the off-center procedure led to a reduction in the frequency aperture from 1890 Hz to 1260 Hz. The expected reduction in 20-Hz operation would be an improvement from 2520 Hz to 1680 Hz.

Radial Off-Center Slip-Stacking debuted at the start of the FY22 run and remained part of Recycler operations. Historically, the final 630-Hz acceleration step in on-center slip-stacking produced DC beam in the Recycler that contributed heavily to losses in the Main Injector. Thus, the implementation of Radial Off-Center Slip-Stacking reduced MI losses by a factor of two [4], significantly improving the efficiency of the accelerator complex.

## MAIN INJECTOR RAMP

To meet 850-kW operations, the Main Injector ramped to 120 GeV every 1.2 s at $5 \times 10^{13}$ protons on target. The nominal intensity in the Recycler during this mode was $5 \times 10^{10}$ protons per bunch (ppb). With the completion of the PIP-II infrastructure, the switch to the 20-Hz cycle rate, and the subsequent increase in beam intensity, the expected beam power will reach 1.25 MW.

Intensity-based strategies for increasing beam power impose risks to the accelerator complex that arise from both systemic and physics-related sources. These concerns warranted investigation during the development of existing operating plans, and efforts to further increase beam power, such as ACE, would need to overcome instabilities and limitations related to pulse intensity if that were the focus. Significant efforts at Fermilab have already explored the impacts of electron clouds in the Recycler [5], TMCI and convective instabilities [6], and their relations to beam intensity. Limited physical vacuum apertures and the ages of components in the accelerator complex introduce some of the systemic issues.

Reducing the MI ramp time yields a strategy for achieving a multi-megawatt facility with far fewer systemic risks and instability challenges—as upgrades are confined to a single machine, and intensity effects generally involve the Booster and Recycler.

The minimum possible MI ramp time that still facilitates slip-stacking in the Recycler is 0.65 s. Achieving this upgrade would require improvements to the power-delivery infrastructure, RF systems, instrumentation and controls—along with a renewed understanding of the MI magnets and their limits. These magnets were designed to ramp at 240 GeV/c/s, but the 0.65-s ramp necessitates $\dot{P}$ = 500 GeV/c/s. We are investigating concerns related to degradation, failure rates, and cooling that could arise due to the rate increase.

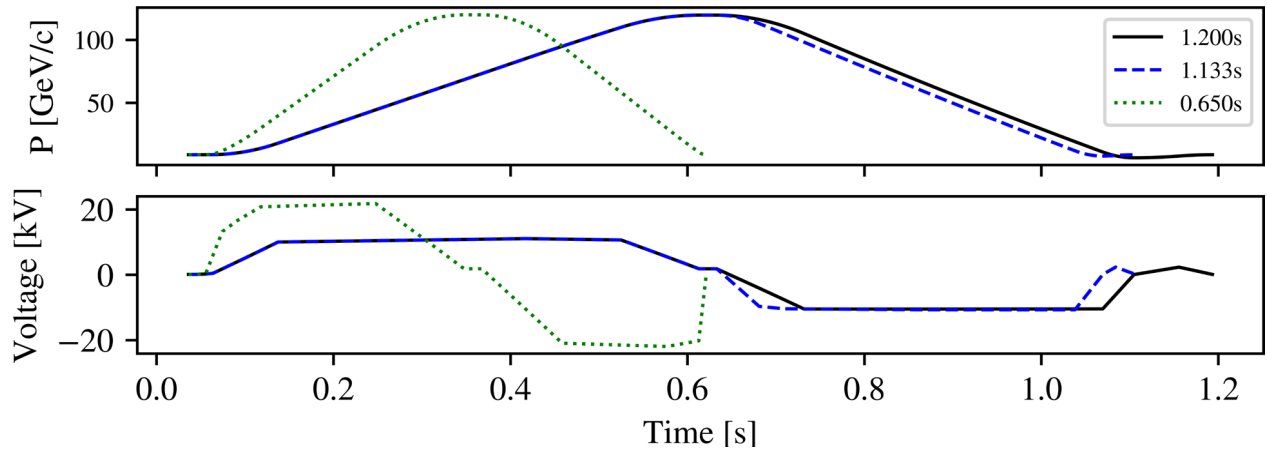

Figure 3: Momentum (*top*) and magnet supply voltage (*bottom*) given a 1.2-s (*solid*), 1.133-s (*dashed*), and 0.65-s (*dotted*) MI ramp time. The 1.2-s case represents nominal running. A 1.133-s ramp was established during FY23, and the 0.65-s ramp is the optimal ACE consideration.

Figure 3 depicts the momentum and magnet supply voltage ramps for three scenarios and demonstrates the increased voltage needed to drive the quickest acceleration. Table 1 summarizes the nominal and anticipated beam powers with the PIP-II and ACE specifications. Note that the response between realized and predicted beam powers is nonlinear due to underlying shifts in the beam intensity and cycle rate.

Table 1: MI beam power capabilities in four different scenarios, including present operations, PIP-II specifications, and ACE-initiative considerations.

| Scenario | Present | PIP-II | ACE-A | ACE-B |
|---|---|---|---|---|
| Ramp Time(s) | 1.2 | 1.2 | 0.9 | 0.7 |
| Ideal $f_b$(Hz) | 15 | 20 | 20 | 20 |
| **MI Power Megawatts** | 0.87 | 1.25 | 1.67 | 2.14 |

To fully realize the ACE plan, we intend to upgrade identified areas through a series of projects between 2024 and 2032. The MI magnet power supplies, abort lines, number of evaporative cooling towers, power substation, electrical feeders, number of RF cavities, and corrector power supplies will need to be addressed. It is estimated that a 0.65-s ramp would deliver 40% percent more protons on target after five years and 68% after ten years with respect to the PIP-II specification [7].

During FY23, we shortened the MI ramp to 1.133 s to demonstrate the feasibility and impact of the time-focused beam power strategy. These endeavours were pursued without changing the Booster cycle rate, which held at an average 14 Hz throughout this part of the run. We achieved a new sustained power record in the MI of 959 kW over one hour during this study, rectified several controls and regulator issues, and prepared to further reduce the ramp duration to 1.067 s. Pending cooling modifications to electrical equipment, the 1.067-s test will occur in FY24, bringing the beam power to around 1.02 MW.

## CONCLUSIONS

The introduction of Radial Off-Center Slip-Stacking significantly improved the efficiency of the Fermilab accelerator complex and reduced beam losses, facilitating increased beam power. We have found that intensity-based approaches for increasing the neutrino flux for LBNF/DUNE can introduce instabilities and exploit systemic vulnerabilities in aging machines. As such, increasing the beam power by reducing the MI ramp duration presents an attractive and viable alternative. A fully realized 0.65-s ramp would turn Fermilab into a multi-megawatt facility that would positively affect DUNE's sensitivity to neutrino physics. Through collaboration with various support divisions at Fermilab, the Main Injector Department set a new sustained power record of 959 kW by marginally decreasing the ramp duration, and a 1.067-s test is planned for FY24 that will usher in the megawatt era.

## ACKNOWLEDGEMENTS

This manuscript has been authored by Fermi Research Alliance, LLC under Contract No. DE-AC02-07CH11359 with the U.S. Department of Energy, Office of Science, Office of High Energy Physics.